\documentclass[twocolumn,showpacs,superscriptaddress,preprintnumbers,amsmath,amssymb,prc]{revtex4-2}
\usepackage{multirow}
\usepackage{amssymb}
\usepackage{amsmath}
\usepackage{graphicx}
\usepackage[normalem]{ulem}
\usepackage{cancel}
\usepackage{multirow}
\usepackage{CJK}
\usepackage[usenames]{color}
\usepackage[colorlinks,linkcolor=blue,urlcolor=blue,citecolor=blue]{hyperref}

\usepackage{tikz,xcolor,hyperref}

\definecolor{lime}{HTML}{A6CE39}
\DeclareRobustCommand{\orcidicon}{
	\begin{tikzpicture}
	\draw[lime, fill=lime] (0,0) 
	circle [radius=0.16] 
	node[white] {{\fontfamily{qag}\selectfont \tiny ID}};
	\draw[white, fill=white] (-0.0625,0.095) 
	circle [radius=0.007];
	\end{tikzpicture}
	\hspace{-2mm}
}
\foreach \x in {A, ..., Z}{%
	\expandafter\xdef\csname orcid\x\endcsname{\noexpand\href{https://orcid.org/\csname orcidauthor\x\endcsname}{\noexpand\orcidicon}}
}

\usepackage{soul,xcolor}
\setstcolor{red}
\newcommand\redsout{\bgroup\markoverwith{\textcolor{red}{\rule[0.5ex]{2pt}{0.4pt}}}\ULon}

\begin{document}

\title{Femtoscopy between  $\pi$, $K$ and $p$ in different heavy-ion collisions  at $\sqrt{s_{NN}}$ = 39 GeV}

\author{Ting-Ting Wang(王婷婷)}
\affiliation{Key Laboratory of Nuclear Physics and Ion-Beam Application (MOE), Institute of Modern Physics, Fudan University, Shanghai 200433, China}
\affiliation{Shanghai Institute of Applied Physics,  Chinese Academy of Sciences, Shanghai 201800, China}

\author{Yu-Gang Ma(马余刚)\orcidB{}} \thanks{Corresponding author:  mayugang@fudan.edu.cn}
\affiliation{Key Laboratory of Nuclear Physics and Ion-Beam Application (MOE), Institute of Modern Physics, Fudan University, Shanghai 200433, China}
\affiliation{Shanghai Research Center for Theoretical Nuclear Physics， NSFC and Fudan University, Shanghai 200438, China}

\author{Song Zhang(张松)\orcidC{} 
}
\affiliation{Key Laboratory of Nuclear Physics and Ion-Beam Application (MOE), Institute of Modern Physics, Fudan University, Shanghai 200433, China}
\affiliation{Shanghai Research Center for Theoretical Nuclear Physics， NSFC and Fudan University, Shanghai 200438, China}

\begin{CJK*} {UTF8} {gbsn}

\date{\today}

\begin{abstract}
Momentum correlation functions between   $\pi$, $K$ and $p$ are calculated for several heavy-ion collision systems, namely $_{5}^{10}\textrm{B}+_{5}^{10}\textrm{B}$, $_{8}^{16}\textrm{O}+_{8}^{16}\textrm{O}$, $_{20}^{40}\textrm{Ca}+_{20}^{40}\textrm{Ca}$ and $_{79}^{197}\textrm{Au}+_{79}^{197}\textrm{Au}$  in  central collisions as well as $_{79}^{197}\textrm{Au}+_{79}^{197}\textrm{Au}$ collision in different centralities at center of mass energy $\sqrt{s_{NN}}$ = 39 GeV within the framework of  A Multi-Phase Transport (AMPT) model complemented by the Lednick$\acute{y}$ and Lyuboshitz analytical method. 
The results present the centrality and system-size dependence of the momentum correlation functions among pairs of $\pi$, $K$ and $p$, from which  the emission source-size can be  deduced. It is found that the deduced  source sizes 
increase with the decreasing of centrality for Au + Au system or with the increasing of system-size in central collisions with different nuclear size. 
In addition, through the momentum correlation functions of nonidentical particle pairs gated on velocity, the average emission sequence of non-identical particles can be indicated. The results illustrate that in the small relative momentum region, protons are emitted in average earlier than $\pi^+$ and $K^+$, and $K^+$ are emitted averagely earlier than $\pi^+$. Furthermore, it seems that larger interval of the average emission order among them is exhibited for  smaller collision systems. The present study sheds light on the dynamics of light particle emission at RHIC energy.

\end{abstract}


\maketitle

\end{CJK*}

\section{Introduction}
In heavy-ion collisions (HICs), two-particle momentum correlation function also called as the Hanbury-Brown Twiss (HBT) interferometry, is different from the original application in astronomy~\cite{Hanbury1956-1,Hanbury1956-2}, and has been normally utilized to extract space-time information of the emission source and probe the dynamical evolution of nuclear collisions in a wide energy range~\cite{Koonin1977,Lisa2005,Heinz,Goldhaber1960,Bialas2000,Ghetti2000,Boal1990,Ardouin1997,RGhetti2003,HeJJ}. Many studies on the two-particle momentum correlation functions of different kinds of particles in HICs can be also found in literature~\cite{wtt2018,Kotte2005,RGhetti2003,DGourio2000,Poch1987,WGGong1991,YGMa2006,RGhetti2004,LWChen2003,DQFang2016,HuangBS2,ShenLei,WangTT-2022,Fang,Li1-SCPMA,Li2-SCPMA,Pratt,Voloshin1997,Ardouin1999,zzq2014,STAR1,STAR2}, e.g., for mesons, neutrons, protons, light charged particles (LCP) and so on. Multi-variable dependences, including centrality, system-size, total momentum of particle pairs, isospin of the emission source, nuclear symmetry energy, nuclear equation of state (EOS) as well as in-medium nucleon-nucleon cross section (NNCS) etc,  of different particle momentum correlation functions contain a wealth of information about the space-time characteristics of HICs. 
Even in antimatter zone, the interaction between antiprotons has been measured with the momentum correlation functions and the Charge-Parity-Time equality of interactions between $p$-$p$ and $\bar{p}$-$\bar{p}$ has been confirmed~\cite{Star-nature}. The momentum correlation functions for  particles with strangeness  have been also discussed, 
for instance $\Lambda$ pairs~\cite{Star-prl}, proton-$\Omega$ and proton-$\Xi$ etc \cite{Star-pOmega,Alice-Nature}. Recently this method is extended to investigate the three-body interaction of hadrons in relativistic heavy-ion collisions~\cite{Alice-3body-2023}. Furthermore, theoretical study has been extended to different kinds of nonidentical particle pairs~\cite{Lednicky1996,wtt2019,YijieWang2022,XiBS,Voloshin1997,Ardouin1999,DGourio2000,RGhetti2003,Kotte1999,wtt2023}, from which  information about emission sequence  between pairs could be extracted as proposed in Ref.~\cite{Gelderloos1994}.

In this work we study on the momentum correlation functions of like-sign and unlike-sign particles between $\pi$, $K$ and $p$ in several ultra-relativistic heavy-ion collisions systems, namely $_{5}^{10}\textrm{B}+_{5}^{10}\textrm{B}$, $_{8}^{16}\textrm{O}+_{8}^{16}\textrm{O}$, $_{20}^{40}\textrm{Ca}+_{20}^{40}\textrm{Ca}$ and $_{79}^{197}\textrm{Au}+_{79}^{197}\textrm{Au}$  in central collisions as well as $_{79}^{197}\textrm{Au}+_{79}^{197}\textrm{Au}$ in different centrailities at $\sqrt{s_{NN}}$ = 39 GeV which were simulated by  A Multi-Phase Transport (AMPT) model~\cite{ZWLin2005,ZWLin-new}. The specific collision energy at 39 GeV was selected because it might be close to the critical point~\cite{PhysRevLett.114.142301, PhysRevLett.130.202301}. On the other hand, the small system physics was very interesting because the similar phenomena have been observed at high multiplicity events in p+p, p+Pb, d+Au collisions in comparison with Au + Au collision~\cite{Nagle2018}, and the system scan was motivated  to probe the medium transport properties~\cite{ZHANG2020135366,PhysRevC.101.021901,Citron,WangYZ}. And $_{8}^{16}\textrm{O}+_{8}^{16}\textrm{O}$ collisions were conducted by STAR collaboration~\cite{STAR-QM2023} and will be performed by ALICE collaboration~\cite{Citron}. Different gating conditions such as centrality, system-size, total particle pair momentum ($P_{tot}$) as well as velocity of particle are applied to explore the momentum correlation functions of particle pairs, namely $\pi$, $K$ and $p$. In particular, we report on the indication of  the emission chronology of the above mesons and baryon, which can be deduced from their corresponding momentum correlation functions gated with velocity of particles in HICs at $\sqrt{s_{NN}}$ = 39 GeV.

The rest of this article is organized as follows. In Section II A, we briefly describe A Multi-Phase Transport  model~\cite{ZWLin2005,ZWLin-new}, then introduce how to calculate the momentum correlation functions of particle pairs by using the Lednick$\acute{y}$  and Lyuboshitz analytical formalism~\cite{Koonin1977,Lednicky2007,lednicky2006,lednicky2009,lednicky2008} in Section II B. In Section III, we summarize the simulated results of the particle momentum correlation functions gated on various parameters in different heavy-ion collisions. Section III A compares the results of like-sign and unlike-sign particle pairs momentum correlation functions with experimental data from the RHIC-STAR collaboration. From Section III B to III D, identical and nonidentical particle momentum correlation functions gated on different conditions are systematically investigated and emission chronology of $\pi$, $K$ and $p$ is  discussed. Finally, a summary is given in Section IV.

\section{MODELS AND FORMALISM}
\subsection{AMPT model}

To obtain phase-space distributions of (anti)particles, A Multi-Phase Transport  model~\cite{ZWLin2005,ZWLin-new} is used as an event generator, which has been applied successfully for studying heavy-ion collisions at relativistic energies~\cite{SZhang2010,Alver2010,GLMa1,LXHan2011,GLMa2,zzq2014,SZhang2017,YiLinCheng2021,ZhangH2021,WangH21,WangH22}. In the version of melting AMPT, the initial phase-space information of partons is generated by the heavy-ion jet interaction generator (HIJING) model~\cite{XNWang1991,MGyulassy1994}. The interaction between partons is then simulated by Zhang's parton cascade (ZPC) model~\cite{BZhang1998}. During the hadronization process, a quark coalescence model is used to combine partons into hadrons~\cite{ZWLin2001,Subrata2004,Subrata2002}. Then, the hadronic rescattering evolution is described by a relativistic transport (ART) model~\cite{BALi1995}. 

The phase-space distributions of particles are  selected at the final stage in the hadronic rescattering process (ART model~\cite{BALi1995}) with considering baryon-baryon, baryon-meson, and meson-meson elastic and inelastic scatterings, as well as resonance decay or week decay. 

\subsection{Lednik{$\acute{y}$}  and Lyuboshitz technique}

AMPT model is not able to  directly give the two-particle momentum correlation function because the quantum statistics (QS) effect and final-state interactions (FSI)~\cite{Koonin1977,Lednicky2007} are not implemented in the model, therefore a momentum correlation function technique has to be introduced. One of the techniques of the two-particle momentum correlation function was proposed  by Lednick$\acute{y}$ and Lyuboshitz~\cite{lednicky2006,lednicky2009,lednicky2008}, which coupled with the phase-space from transport model or a simple Gaussian emission source to produce momentum correlation function. The method  is based on the principle as follows: when two particles emitted at small relative momentum, their momentum correlation function is determined by the space-time characteristics of the production processes owing to the effects of QS and FSI~\cite{Koonin1977,Lednicky2007}.

According to the conditions in Ref.~\cite{Lednicky1996}, we obtain the following expression:
\begin{equation}
\textbf{C}\left(\textbf{k}^*\right) = \frac{\int
\textbf{S}\left(\textbf{r}^*,\textbf{k}^*\right)
\left|\Psi_{\textbf{k}^*}\left(\textbf{r}^*\right)\right|^{2}d^{4}\textbf{r}^*}
{\int
\textbf{S}\left(\textbf{r}^*,\textbf{k}^*\right)d^{4}\textbf{r}^*},
\end{equation}
where $\textbf{r}^* = \textbf{x}_{1}-\textbf{x}_{2}$ and $\textbf{k}^* = \textbf{q} = \frac{1}{2}(\textbf{p}_{1}-\textbf{p}_{2})$ are the relative distance and half of the relative momentum of the two particles in the pair rest frame (PRF) at their kinetic freeze-out, respectively. $\textbf{S}\left(\textbf{r}^*,\textbf{k}^*\right)$ is the probability to emit a particle pair with given $\textbf{r}^*$ and $\textbf{k}^*$, $i.e.$, the source emission function, and $\Psi_{\textbf{k}^*}\left(\textbf{r}^*\right)$ is the equal-time $\left(t^* = 0\right)$ reduced Bethe-Salpeter amplitude which can be approximated by the outer solution of the scattering problem in the PRF. 
 
Momentum correlations of both identical and non-identical particle pairs are influenced by the strong and Coulomb-induced correlations. 
In this calculation, for non-identical charged pair, Coulomb interaction is dominant. Strong interaction is also present, but is expected to be small. The reduced Bethe-Salpeter amplitude can be approximated by the outer solution of the scattering problem. This is
\begin{multline}
\Psi_{\textbf{k}^*}\left(\textbf{r}^*\right) = e^{i\delta_{c}}\sqrt{A_{c}\left(\lambda \right)} \times\\
\left[e^{-i\textbf{k}^*\textbf{r}^*}F\left(-i\lambda,1,i\xi\right)+f_c\left(k^*\right)\frac{\tilde{G}\left(\rho,\lambda \right)}{r^*}\right],
\end{multline}
where $\delta_{c} =$ $arg$ $\Gamma\left(1+i\lambda\right)$ is the Coulomb $s$-wave phase shift with $\lambda = \left(k^*a_c\right)^{-1}$. Here $a_{c}$ is the two-particle Bohr radius, which is equal to 387 fm for pion-pion, and equal to 248.52, 222.56, and 83.59 fm for pion-kaon, pion-proton, and kaon-proton pairs. $A_c\left(\lambda \right) = 2\pi\lambda \left[\exp\left(2\pi\lambda \right)-1\right]^{-1}$ is the Coulomb penetration factor, and its positive (negative) value corresponds to the repulsion (attraction). $\tilde{G}\left(\rho,\lambda \right) = \sqrt{A_{c}\left(\lambda \right)}\left[G_0\left(\rho,\lambda \right)+iF_0\left(\rho,\lambda \right)\right]$ is a combination of regular $\left(F_0\right)$ and singular $\left(G_0\right)$ $s$-wave Coulomb functions~\cite{lednicky2009,lednicky2008}. $F\left(-i\lambda,1,i\xi\right) = 1+\left(-i\lambda\right)\left(i\xi\right)/1!^{2}+\left(-i\lambda\right)\left(-i\lambda+1\right)\left(i\xi\right)^{2}/2!^{2}+\cdots$ is the confluent hypergeometric function with $\xi = \textbf{k}^*\textbf{r}^*+\rho$, $\rho = k^*r^*$.
\begin{equation}
f_c\left(k^*\right) = \left[ K_{c}\left(k^*\right)-\frac{2}{a_c}h\left(\lambda \right)-ik^*A_{c}\left(\lambda \right)\right]^{-1}
\end{equation}
is the $s$-wave scattering amplitude renormalizied by the long-range Coulomb interaction, with $h\left(\lambda \right) = \lambda^{2}\sum_{n=1}^{\infty}\left[n\left(n^2+\lambda^2\right)\right]^{-1}-C-\ln\left[\lambda \right]$ where $C$ = 0.5772 is the Euler constant. 
$K_{c}\left(k^*\right) = \frac{1}{f_0} + \frac{1}{2}d_0k^{*^2} + Pk^{*^4} + \cdots$ is the effective range function, where $d_{0}$ is the effective radius of the strong interaction, $f_{0}$ is the scattering length and $P$ is the shape parameter. The parameters of the effective range function are important parameters characterizing the essential properties of the FSI. The discrepancy for different particle pairs can also influence the effect of FSI on source size.

The details on the formalism of the two-particle momentum correlation function can be also found in Ref.~\cite{wtt2019,wtt2023}.

\section{ANALYSIS AND DISCUSSION}
 
\subsection{Comparison of correlation functions between like-sign and unlike-sign particle pairs   with experimental data}

\begin{figure}[!htbp]
 \includegraphics[width=1.1\linewidth]{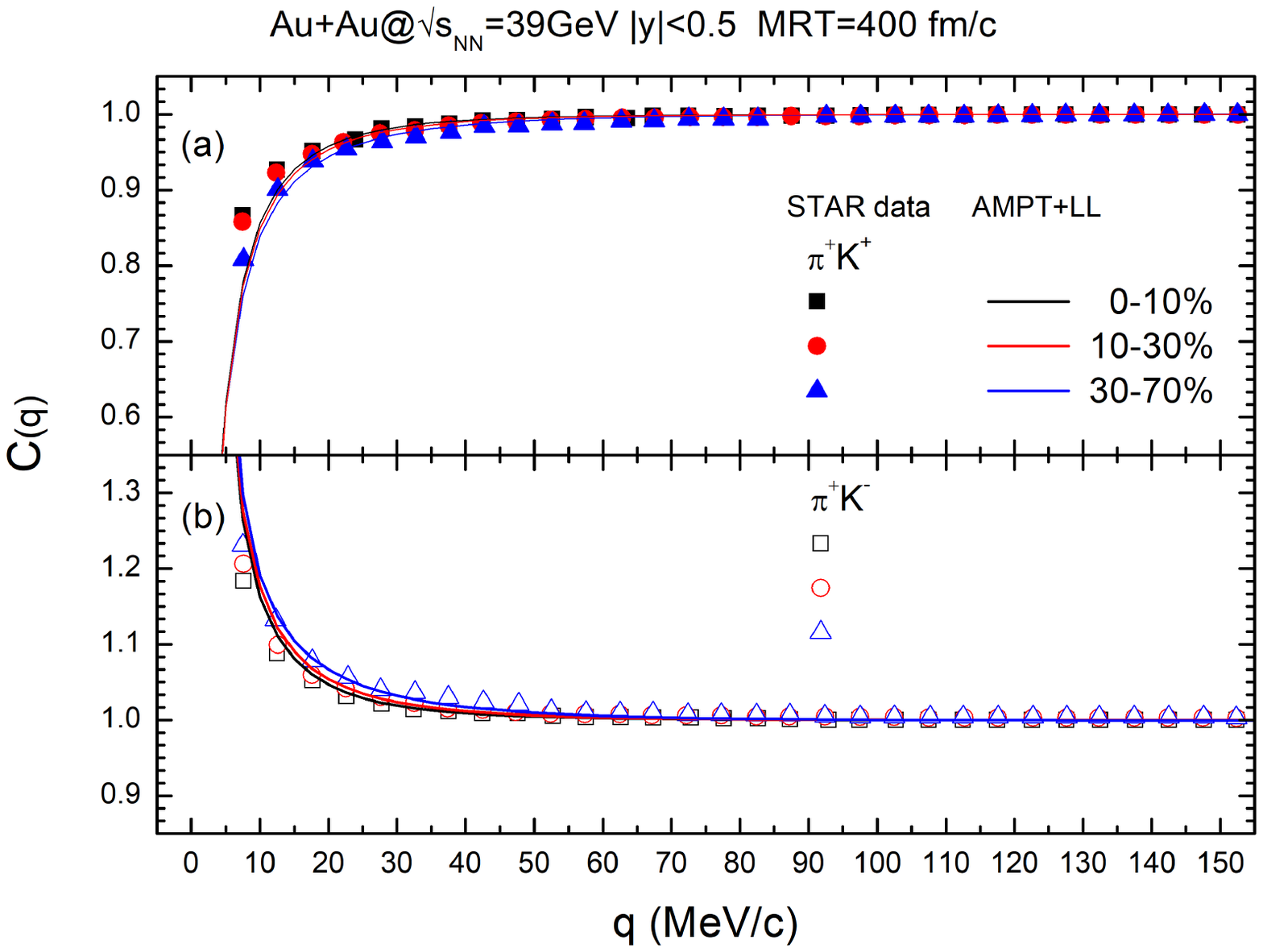}
 \centering
 \caption{Momentum correlation functions for like-sign (a) and unlike-sign (b) $\pi$-$K$ pairs for Au+Au collisions at $\sqrt{s_{NN}}$ = 39 GeV in different centrality classes ($0-10$ $\%$, $10-30$ $\%$, and $30-70$ $\%$). Solid markers represent the preliminary experimental data from the RHIC-STAR collaboration~\cite{Zbroszczyk2019,Siejka2019}. Lines represent theoretical fits calculated by  the AMPT model plus the Lednick$\acute{y}$ and Lyuboshitz code. Note that the longer hadronic rescattering time of 400 $fm/c$ for this meson-meson pair is used in this specific calculation for comparing with the data.}
 \label{fig1}
\end{figure}

\begin{figure}[!htbp]
 \includegraphics[width=\linewidth]{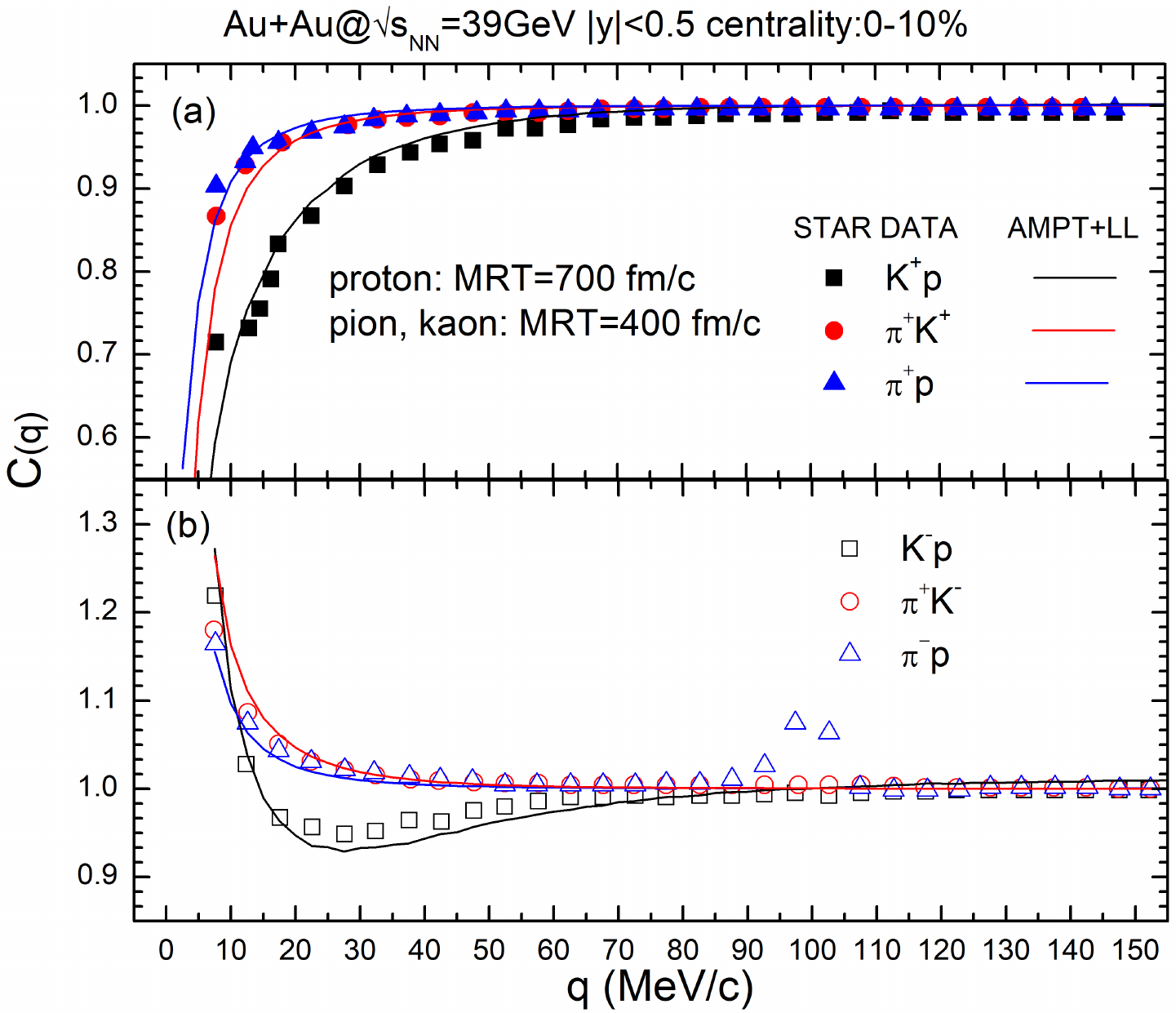}
 \centering
 \caption{ Momentum correlation functions for  like-sign (a) and unlike-sign (b) for  kaon-proton, pion-kaon, and pion-proton pairs for central ($0-10$ $\%$) Au + Au collisions at $\sqrt{s_{NN}}$ = 39 GeV. Solid markers represent the preliminary experimental data from the RHIC-STAR collaboration~\cite{Zbroszczyk2019,Siejka2019}. Lines represent theoretical fits calculated using the AMPT model plus the Lednick$\acute{y}$ and Lyuboshitz code. Note that the longer hadronic rescattering time of 400 $fm/c$ for meson-meson pairs and 700 $fm/c$ for proton-meson pair are used in this specific calculation for comparing with the data.
 }
 \label{fig2}
\end{figure}

Fig.~\ref{fig1} presents results of $\pi^+$-$K^+$ (a) and $\pi^+$-$K^-$ (b) correlation functions for three different centrality classes of $0-10$ $\%$, $10-30$ $\%$, and $30-70$ $\%$ calculated by the AMPT model combined with Lednick$\acute{y}$ and Lyuboshitz code in Au + Au collisions at $\sqrt{s_{NN}}$ = 39 GeV. Within the cut of transverse momentum $p_{t}$ and rapidity $y$, we confront the experimental data with the predictions of the above analysis method. When the phase-space information of mesons with the maximum rescattering time (MRT) of 400 $fm/c$ is selected from the AMPT model, it is found that the results can well describe the experimental data for like-sign (unlike-sign) $\pi$-$K$ momentum correlation functions from the RHIC-STAR collaboration~\cite{Zbroszczyk2019,Siejka2019}. We can also observe a slight centrality dependence of  momentum correlation function for meson pairs.  Through the same conditions and calculations as Fig.~\ref{fig1}, it is found that the results of Fig.~\ref{fig2} can also well describe the experimental data for the momentum correlation functions of like-sign (unlike-sign) $K$-$p$, $\pi$-$K$, $\pi$-$p$ from the RHIC-STAR collaboration~\cite{Zbroszczyk2019,Siejka2019}. In this case, we fixed the MRT at 700 fm/c for pairs of particles containing proton~\cite{wtt2023}, which could be a reasonable choice for making quantitative comparison with experimental data. However, the quantitative reproduction is not our main concern in the present work. 
We have checked some results for $K$-$p$, $\pi$-$K$, $\pi$-$p$ correlations with different MRT, such as  100 fm/c, 400 fm/c and 700 fm/c, and find that 
 the momentum correlations of above particles are insensitive to MRT, thus the MRT = 100 fm/c is basically safe choice in the present work. Similar procedure for light nuclei correlation has been performed in our previous work for the same systems at $\sqrt{s_{NN}}$ = 39 GeV \cite{wtt2023}.
In order to ensure the statistics to reduce the error, we fixed the MRT at 100 fm/c in the following calculations.

\subsection{Centrality, system-size and $P_{tot}$ dependence of like-sign identical meson pairs momentum correlation functions }

\begin{figure}[!htbp]
 \includegraphics[width=\linewidth]{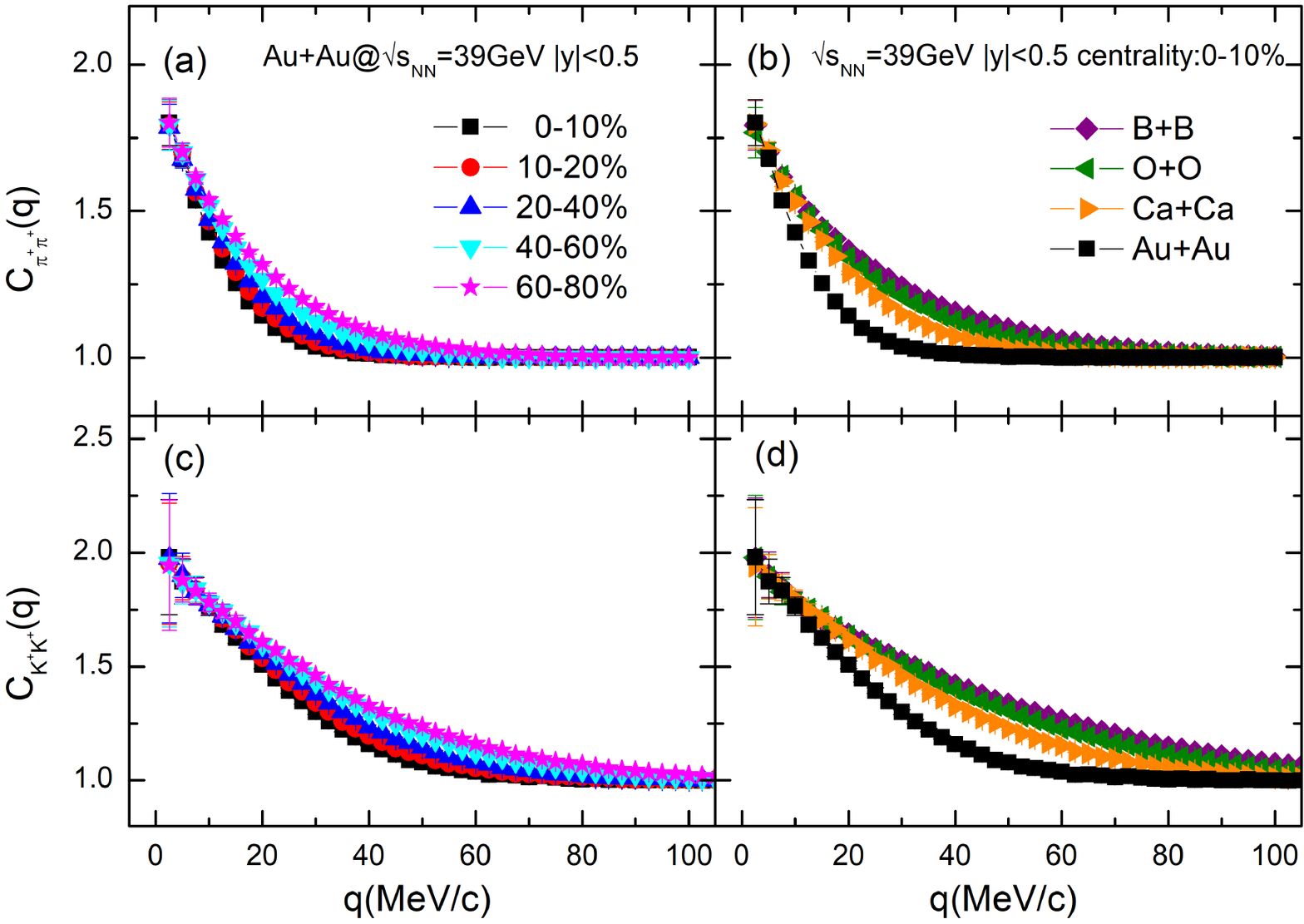}
 \centering
 \caption{Momentum correlation functions at mid-rapidity  ($\left|y \right|< 0.5$)  of pion-pairs and koan-pairs as a function of five different centralities for $_{79}^{197}\textrm{Au}+_{79}^{197}\textrm{Au}$ collisions at $\sqrt{s_{NN}}$ = 39 GeV are presented in (a) and (c), respectively. Momentum correlation functions of pion-pairs and koan-pairs at mid-rapidity ($\left|y \right|< 0.5$) for 0-10$\%$ central collisions of $_{5}^{10}\textrm{B}+_{5}^{10}\textrm{B}$, $_{8}^{16}\textrm{O}+_{8}^{16}\textrm{O}$, $_{20}^{40}\textrm{Ca}+_{20}^{40}\textrm{Ca}$ as well as $_{79}^{197}\textrm{Au}+_{79}^{197}\textrm{Au}$ systems at $\sqrt{s_{NN}}$ = 39 GeV are presented in (b) and (d), respectively. 
 }
 \label{fig3}
\end{figure}

We now perform the systematic analysis on correlation function for like-sign identical meson pairs. The $\pi^+$-$\pi^+$ and $K^+$-$K^+$ correlation functions will be discussed  with specific gates on centrality and system-size. Fig.~\ref{fig3} (a) and (c) present the momentum correlation functions of identical meson pairs for $_{79}^{197}\textrm{Au}+_{79}^{197}\textrm{Au}$ collisions at different centralities of $0-10$ $\%$, $10-20$ $\%$, $20-40$ $\%$, $40-60$ $\%$, and $60-80$ $\%$ at $\sqrt{s_{NN}}$ = 39 GeV. The momentum correlation functions of identical meson pairs exhibit more than unity in Fig.~\ref{fig3}, which is caused by the interplay between the quantum statistical (QS) and final state interactions (FSI), and the shape is consistent with previous results~\cite{zzq2014,Adamczewski2020,Star-prc2005}. 
The enhanced strength of the $\pi^+$-$\pi^+$ and $K^+$-$K^+$ momentum correlation functions is observed in peripheral collisions. These results indicate that meson emission occurs from a source with smaller space  extent in peripheral collision. In addition, the effect of system-size on the momentum correlation functions of mesons is also investigated by four different systems, namely $_{5}^{10}\textrm{B}+_{5}^{10}\textrm{B}$, $_{8}^{16}\textrm{O}+_{8}^{16}\textrm{O}$, $_{20}^{40}\textrm{Ca}+_{20}^{40}\textrm{Ca}$ and $_{79}^{197}\textrm{Au}+_{79}^{197}\textrm{Au}$ in central collisions. In Fig.~\ref{fig3} (b) and (d), the $\pi^+$-$\pi^+$ and $K^+$-$K^+$ momentum correlation functions appear strong sensitivity to system-size and an enhanced strength is observed when meson pairs are emitted from smaller system collisions. This  enhanced strength of the momentum correlation functions for meson pairs is a physical effect stemming from the smaller space extent of the emission source~\cite{Ghetti2000}. Therefore, the emission source-size of meson pairs obtained by their momentum correlation functions and system-size is self-consistent. 

\begin{figure}[!htbp]
 \includegraphics[width=\linewidth]{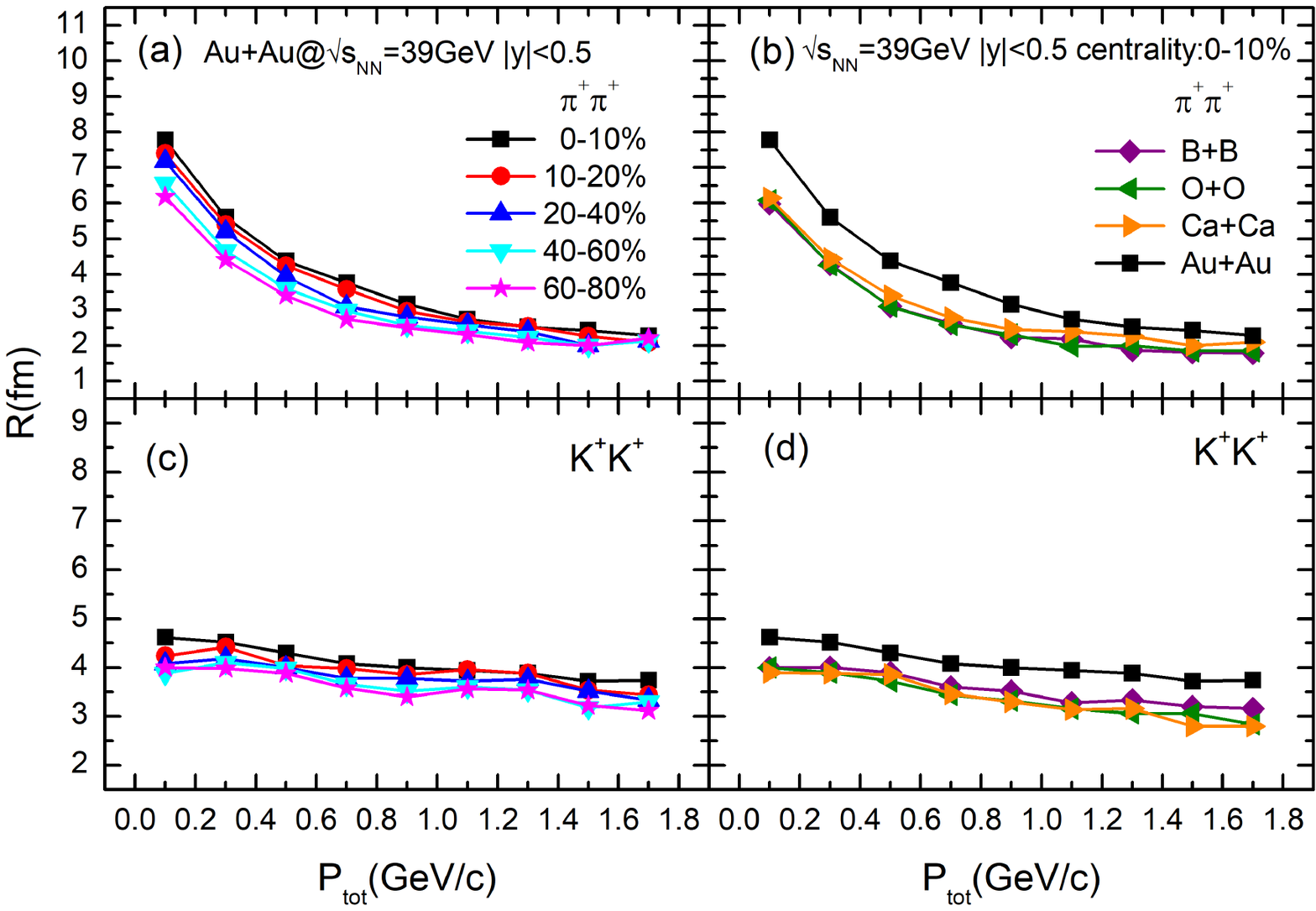}
 \centering
 \caption{$P_{tot}$ dependence of the HBT radii at mid-rapidity  ($\left|y \right|< 0.5$)  of pion-pairs and koan-pairs as a function of five different centralities for $_{79}^{197}\textrm{Au}+_{79}^{197}\textrm{Au}$ collisions at $\sqrt{s_{NN}}$ = 39 GeV are presented in (a) and (c), respectively. $P_{tot}$ dependence of the HBT radii of pion-pairs and koan-pairs at mid-rapidity ($\left|y \right|< 0.5$) for 0-10$\%$ central collisions of $_{5}^{10}\textrm{B}+_{5}^{10}\textrm{B}$, $_{8}^{16}\textrm{O}+_{8}^{16}\textrm{O}$, $_{20}^{40}\textrm{Ca}+_{20}^{40}\textrm{Ca}$ as well as $_{79}^{197}\textrm{Au}+_{79}^{197}\textrm{Au}$ systems at $\sqrt{s_{NN}}$ = 39 GeV are presented in (b) and (d), respectively. 
 }
 \label{fig4}
\end{figure}

In order to observe directly the contributions from centrality and system-size on the two-particle correlations, then the Gaussian source radius is extracted by assuming a Gaussian-type emission source, i.e.,  $S\left(\textbf{r}^*\right)\sim\exp\left(-\textbf{r}^{*^{2}}/\left(4r_{0}^2\right)\right)$, where $r_{0}$ is the Gaussian source radius from the correlation functions. Figure~\ref{fig4} shows the $P_{tot}$ ($P_{tot} = \textbf{p}_{1}+\textbf{p}_{2}$)  dependence of the Gaussian source radius of the two-meson momentum correlation functions for different centrality and system-size by the AMPT model. Figure~\ref{fig4} (a) and (c) present the HBT radius of the two-meson for $_{79}^{197}\textrm{Au}+_{79}^{197}\textrm{Au}$ collisions at different centralities of $0-10$ $\%$, $10-20$ $\%$, $20-40$ $\%$, $40-60$ $\%$, and $60-80$ $\%$ at $\sqrt{s_{NN}}$ = 39 GeV. It is seen from Figure~\ref{fig4} that the radius decreases with the increasing of transverse momentum. The trend of these results is consistent with those from theory and the STAR experiment for the same system~\cite{zzq2014,Star-prc2005}. Qualitatively speaking, high transverse momentum mesons are ejected from the emission source earlier, while the low transverse momentum meson emits later; therefore, we can see the expansion of the emission source by the $P_{tot}$ dependence of the radii.  It does not conflict with the understanding via radial flow effect~\cite{Star-prc2005,Fabrice-PRC2004}, namely the particles with higher $p_t$ may decouple earlier than those with lower $p_t$ from the fireball. Meanwhile, the $P_{tot}$  dependence of the radius of the two-meson momentum correlation functions appear slight sensitivity to centrality and system-size in Fig.~\ref{fig4}. However, the sensitivity seems disappear in the small systems as shown in Fig.~\ref{fig4} (b) and (d). In addition, the $P_{tot}$  dependence of the radius for the $\pi^+$-$\pi^+$ appears more sensitive to centrality and system-size  than the one for $K^+$-$K^+$.  

\subsection{ Momentum correlation functions for like-sign (unlike-sign) nonidentical particles gated on centrality and system-size}

\begin{figure}[!htbp]
 \includegraphics[width=1.0\linewidth]{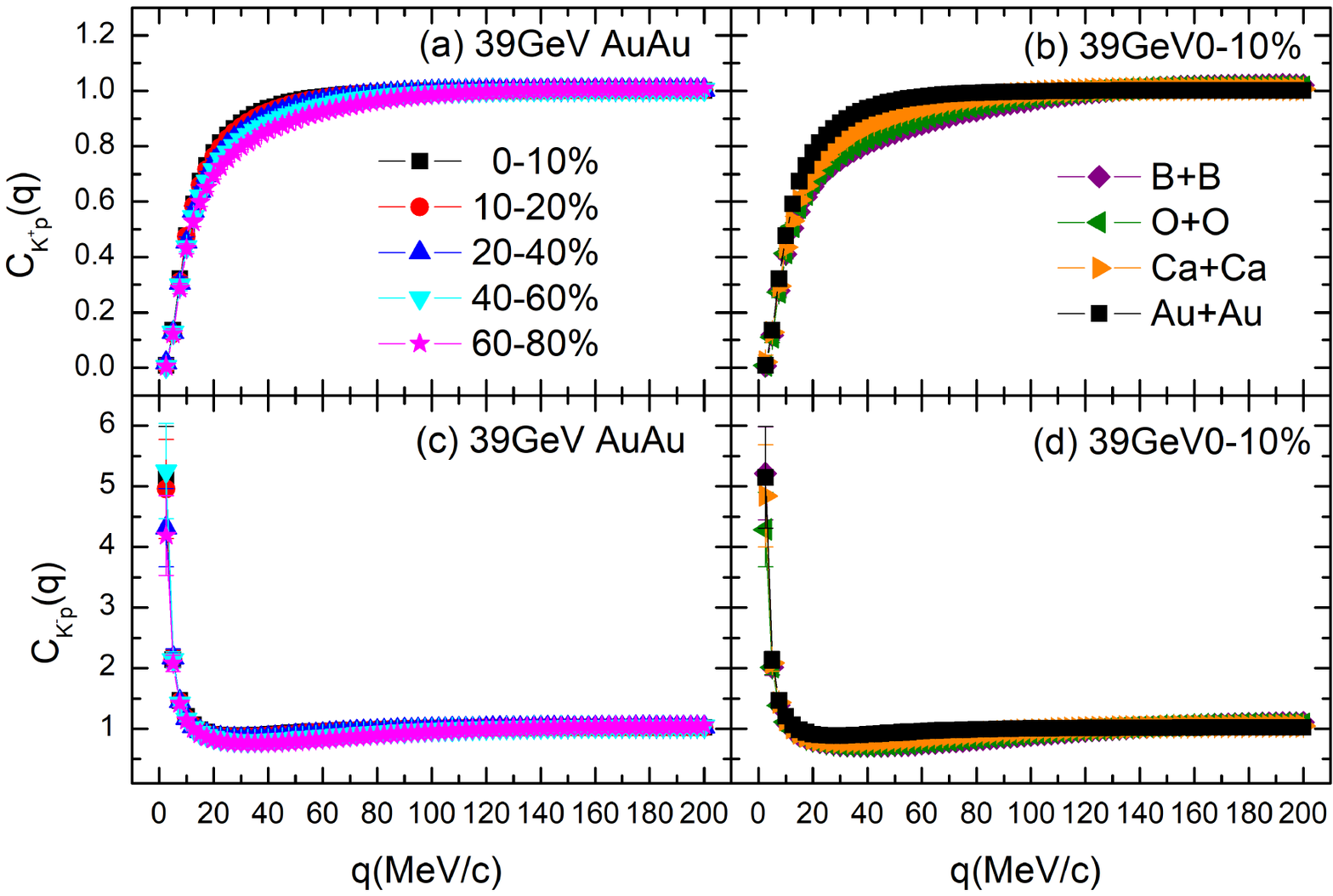}
 \centering
 \caption{Momentum correlation functions at mid-rapidity  ($\left|y \right|< 0.5$)  of like-sign (unlike-sign) kaon-proton as a function of five different centralities for $_{79}^{197}\textrm{Au}+_{79}^{197}\textrm{Au}$ collisions at $\sqrt{s_{NN}}$ = 39 GeV are presented in (a) and (c), respectively. Momentum correlation functions of like-sign (unlike-sign) kaon-proton at mid-rapidity ($\left|y \right|< 0.5$) for 0-10$\%$ central collisions of $_{5}^{10}\textrm{B}+_{5}^{10}\textrm{B}$, $_{8}^{16}\textrm{O}+_{8}^{16}\textrm{O}$, $_{20}^{40}\textrm{Ca}+_{20}^{40}\textrm{Ca}$ as well as $_{79}^{197}\textrm{Au}+_{79}^{197}\textrm{Au}$ systems at $\sqrt{s_{NN}}$ = 39 GeV are presented in (b) and (d), respectively.
 }
 \label{fig5}
\end{figure}

\begin{figure}[!htbp]
 \includegraphics[width=\linewidth]{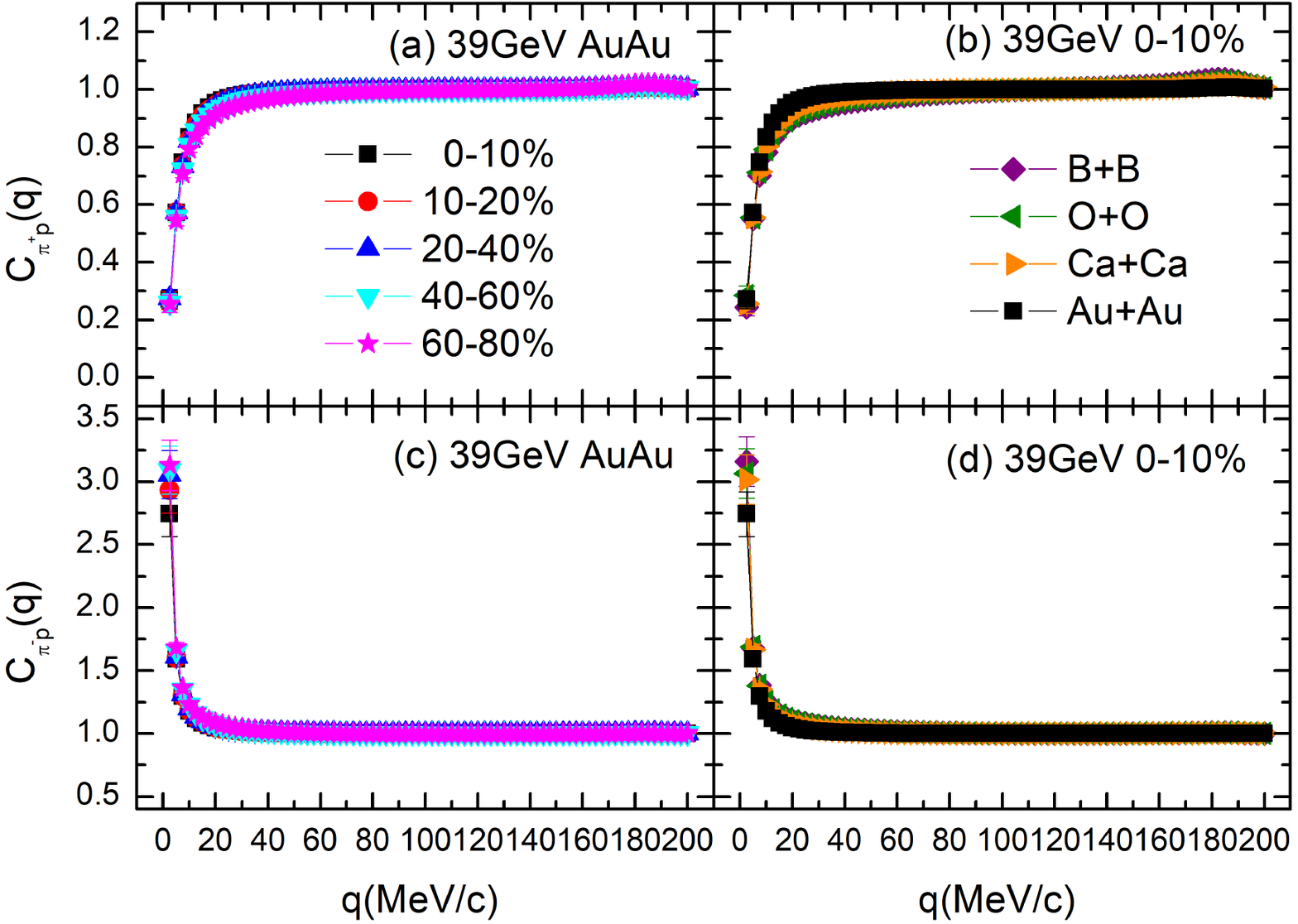}
 \centering
 \caption{Same as Fig.~\ref{fig5} but for pion-proton.
 }
 \label{fig6}
\end{figure}

\begin{figure}[!htbp]
 \includegraphics[width=\linewidth]{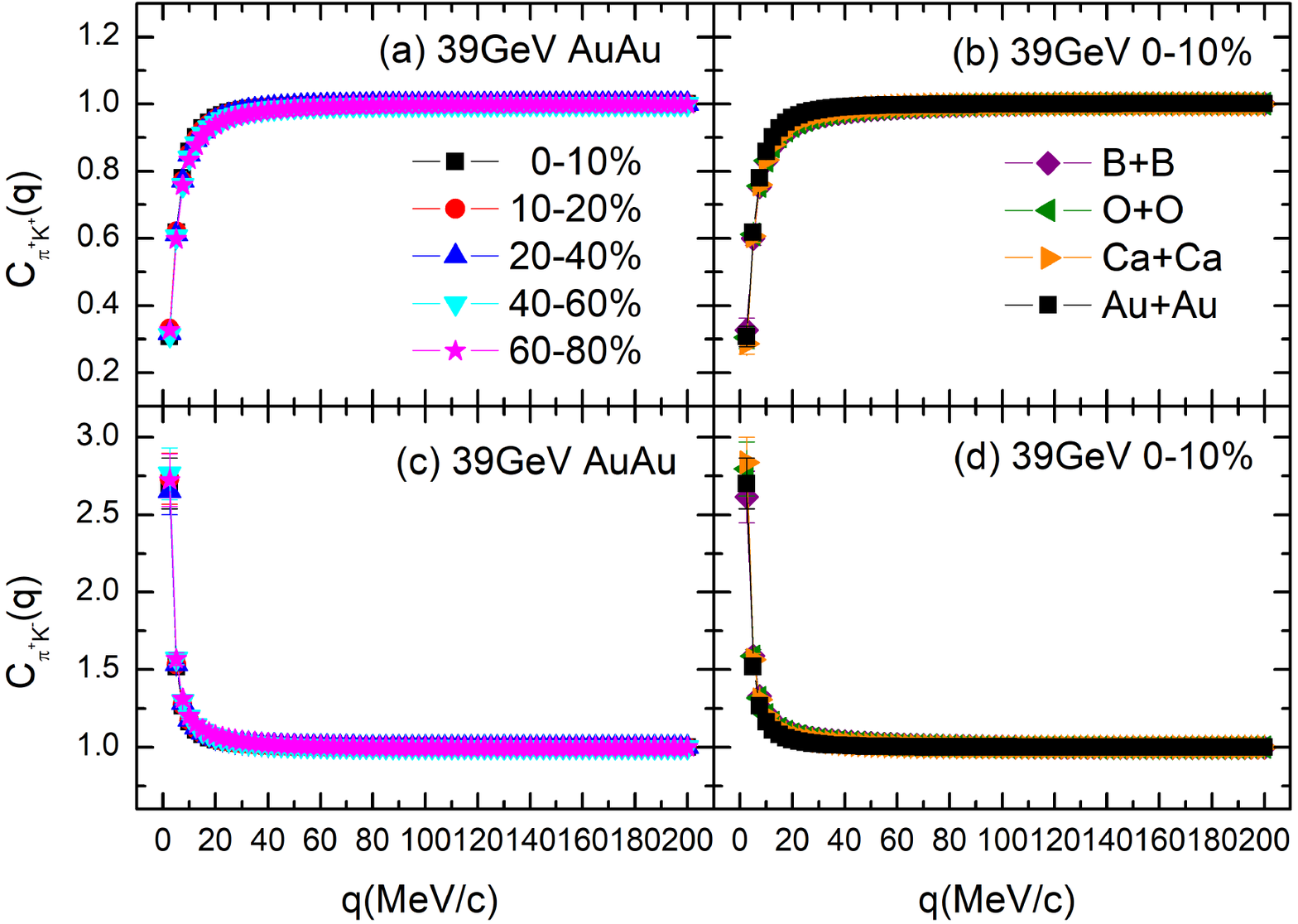}
 \centering
 \caption{Same as Fig.~\ref{fig5} but for pion-kaon. 
 }
 \label{fig7}
\end{figure}

\begin{figure}[!htbp]
 \includegraphics[width=\linewidth]{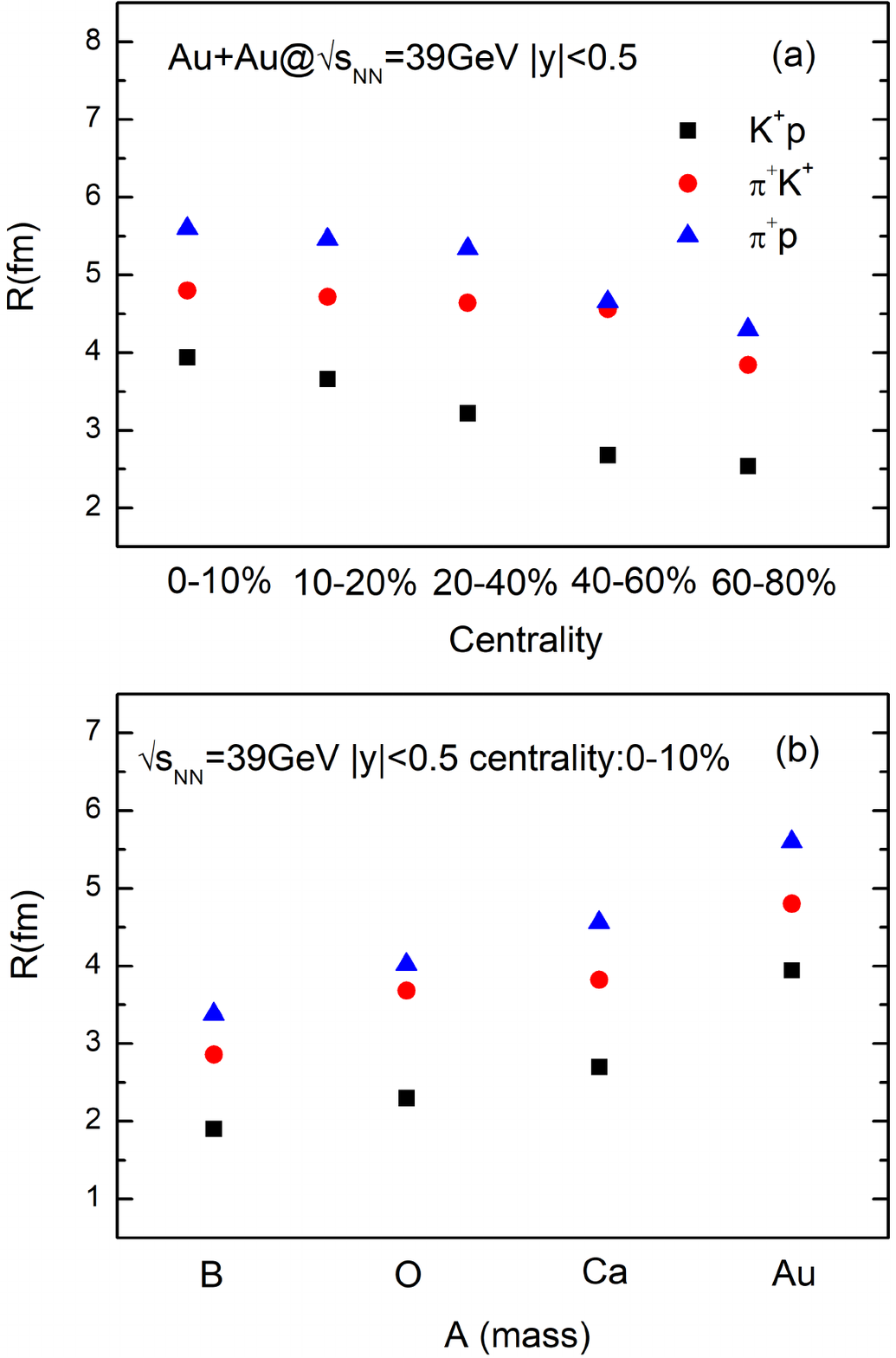}
 \centering
 \caption{(a) Centrality dependence of kaon-proton, pion-kaon and pion-proton source size ($R$) for Au+Au collisions at $\sqrt{s_{NN}}$ = 39 GeV. (b) System dependence of kaon-proton, pion-kaon and pion-proton source size ($R$) for 0-10$\%$ central collisions of $_{5}^{10}\textrm{B}+_{5}^{10}\textrm{B}$, $_{8}^{16}\textrm{O}+_{8}^{16}\textrm{O}$, $_{20}^{40}\textrm{Ca}+_{20}^{40}\textrm{Ca}$ as well as $_{79}^{197}\textrm{Au}+_{79}^{197}\textrm{Au}$ systems at $\sqrt{s_{NN}}$ = 39 GeV.
}
 \label{fig8}
\end{figure}

\begin{figure}[!htbp]
 \includegraphics[width=\linewidth]{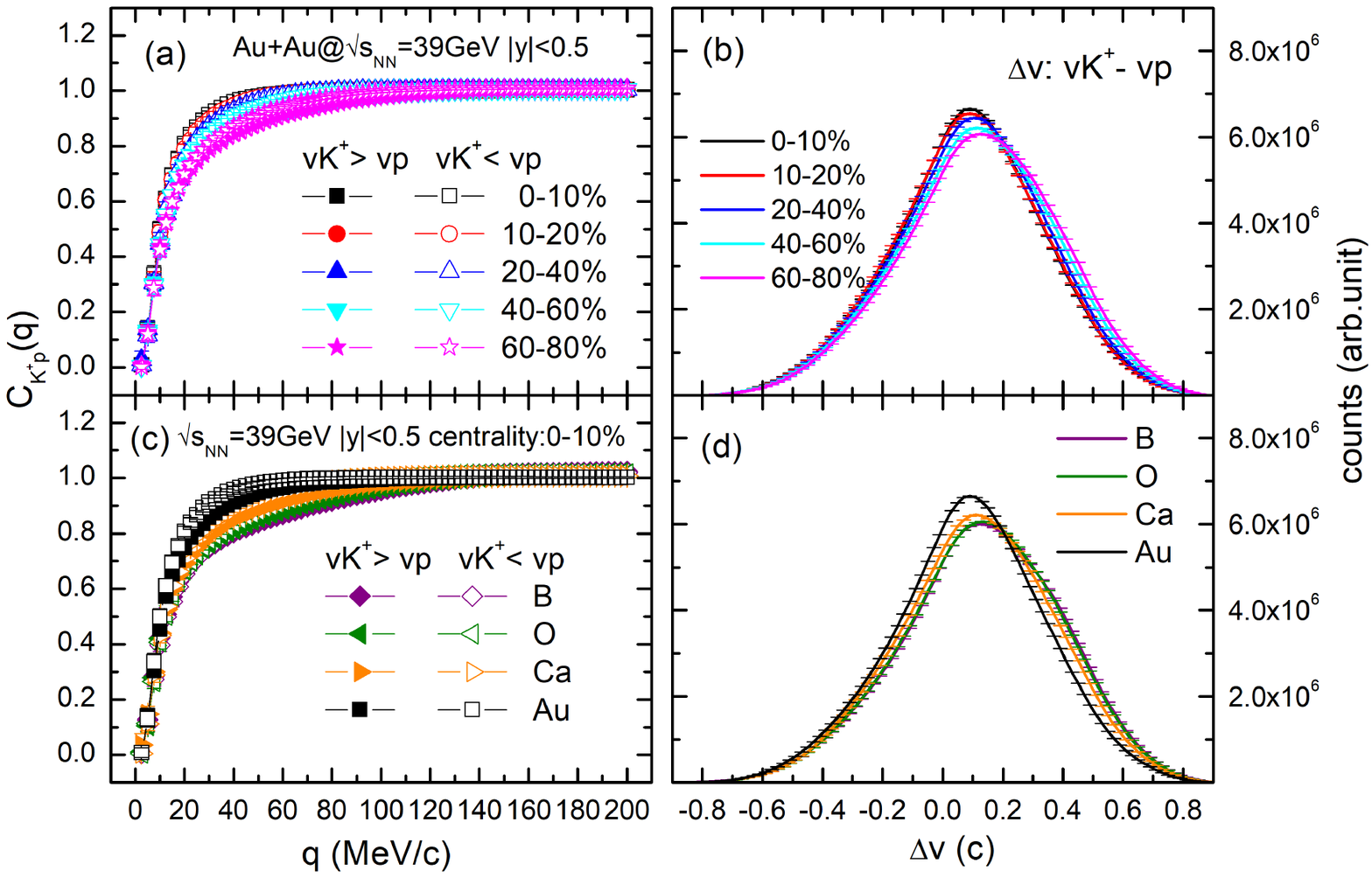}
 \centering
 \caption{Centrality and system dependence of kaon-proton (like-sign) velocity-gated momentum correlation functions and velocity difference ($\Delta v $) by the AMPT + LL model. The velocity conditions are indicated in each panel: $\Delta v >0$ is remarked by solid symbol and the $\Delta v < 0$ by open symbol.}
 \label{fig9}
\end{figure}
\begin{figure}[!htbp]
 \includegraphics[width=\linewidth]{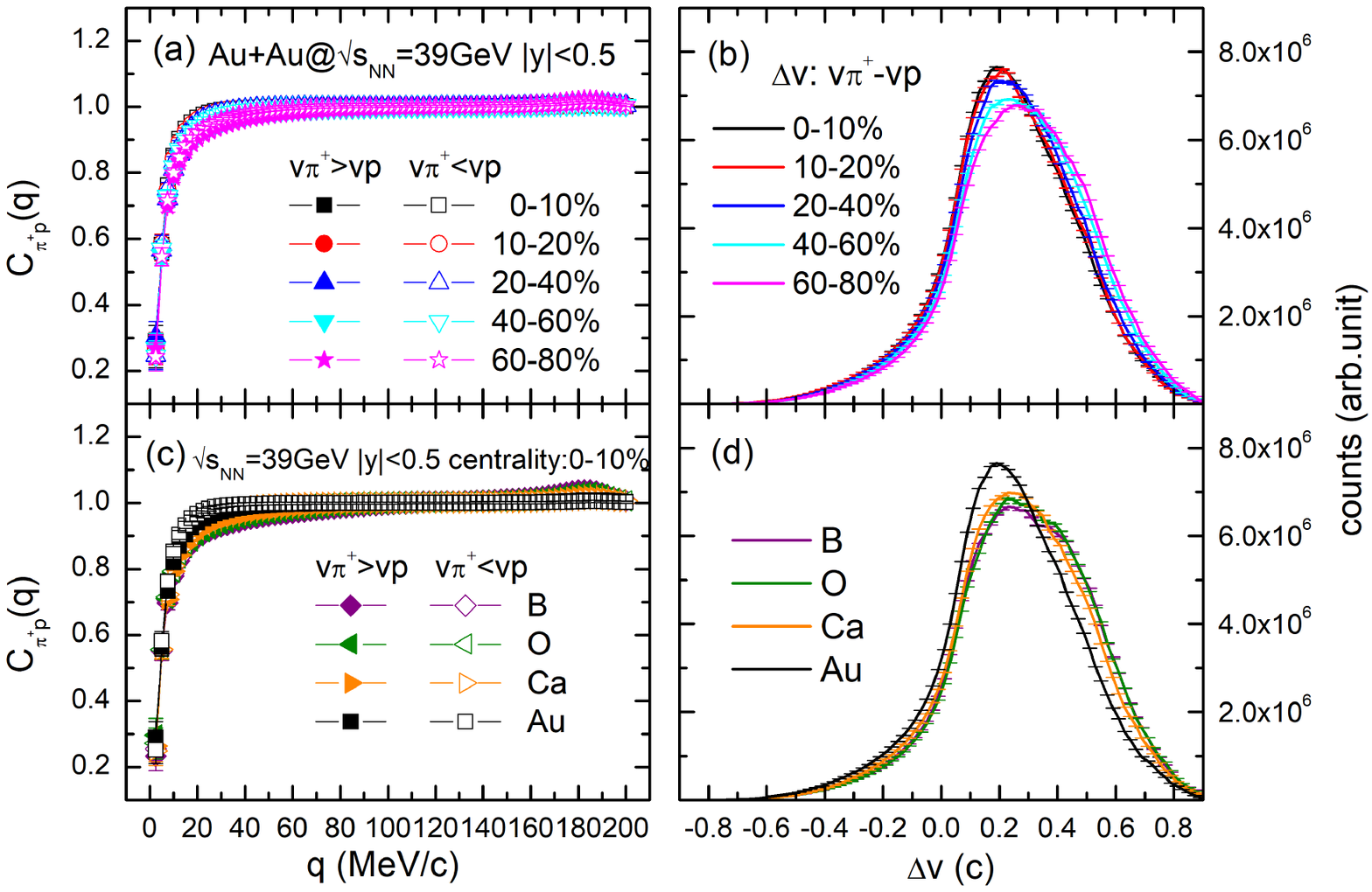}
 \centering
 \caption{Same as Fig.~\ref{fig9} but for pion-proton. 
}
 \label{fig10}
\end{figure}

\begin{figure}[!htbp]
 \includegraphics[width=\linewidth]{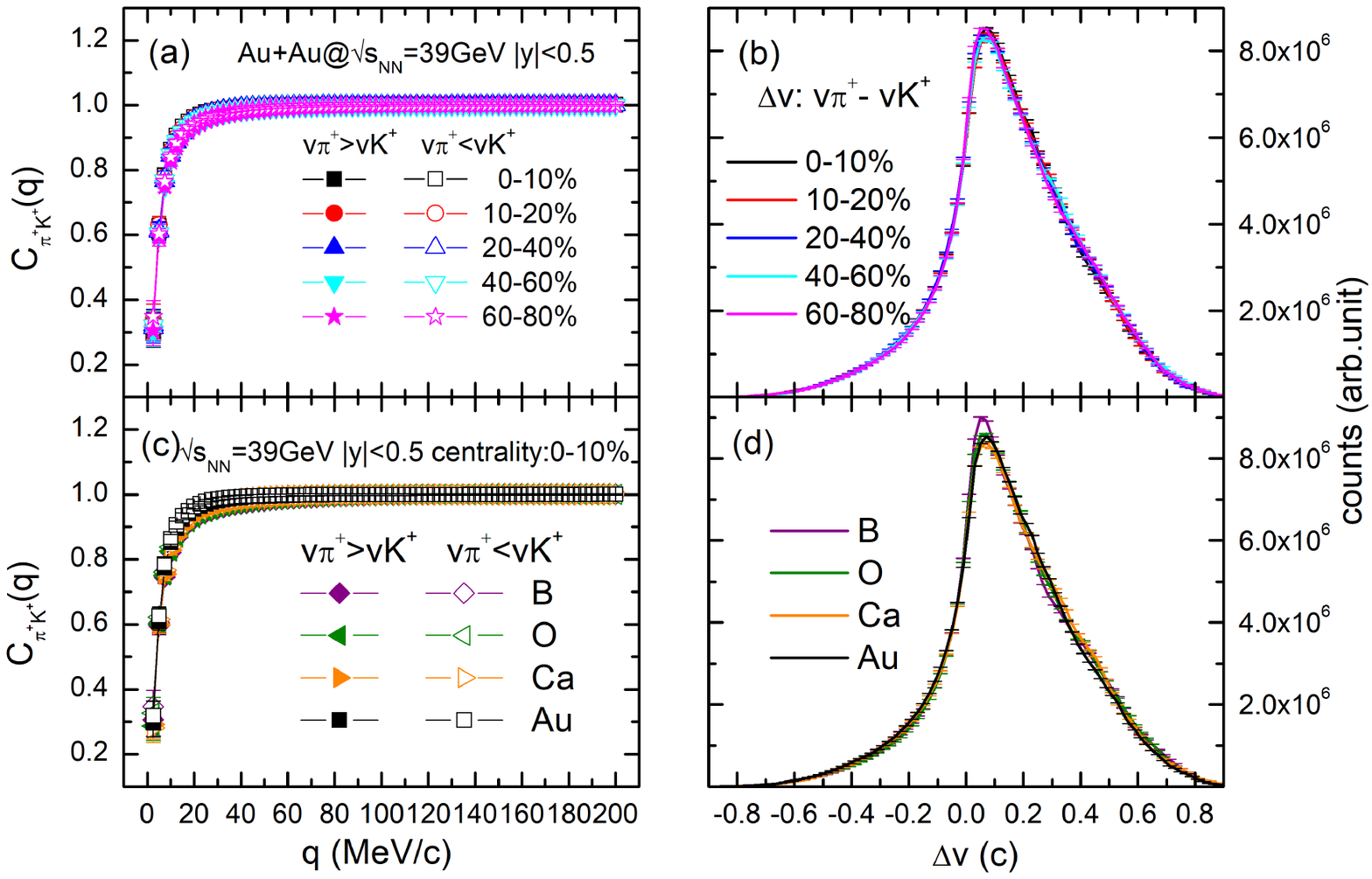}
 \centering
 \caption{Same as Fig.~\ref{fig9} but for pion-kaon. 
 }
 \label{fig11}
\end{figure}


\begin{figure}[!htbp]
 \includegraphics[width=\linewidth]{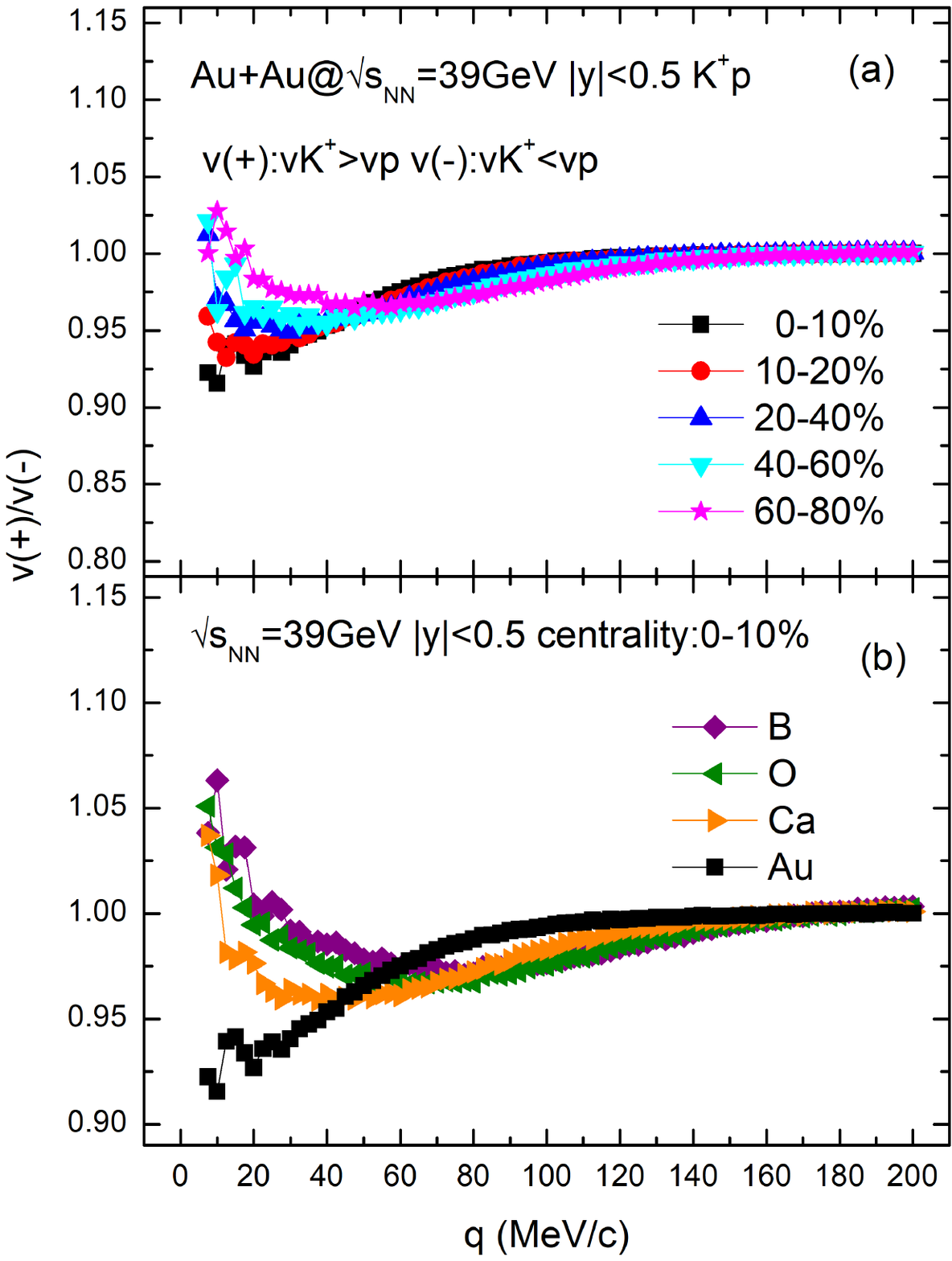}
 \centering
 \caption{(a) Ratios of the velocity-gated momentum correlation functions of kaon-proton for 39 GeV $_{79}^{197}\textrm{Au}+_{79}^{197}\textrm{Au}$ collision at  mid-rapidity ($\left|y \right|< 0.5$)  and five different centralities. (b) Ratios of the velocity-gated momentum correlation functions of kaon-proton for 0 $-$ 10 $\%$ central collisions of $_{5}^{10}\textrm{B}+_{5}^{10}\textrm{B}$, $_{8}^{16}\textrm{O}+_{8}^{16}\textrm{O}$, $_{20}^{40}\textrm{Ca}+_{20}^{40}\textrm{Ca}$ as well as $_{79}^{197}\textrm{Au}+_{79}^{197}\textrm{Au}$ systems at $\sqrt{s_{NN}}$ = 39 GeV.
 }
 \label{fig12}
\end{figure}


\begin{figure}[!htbp]
 \includegraphics[width=\linewidth]{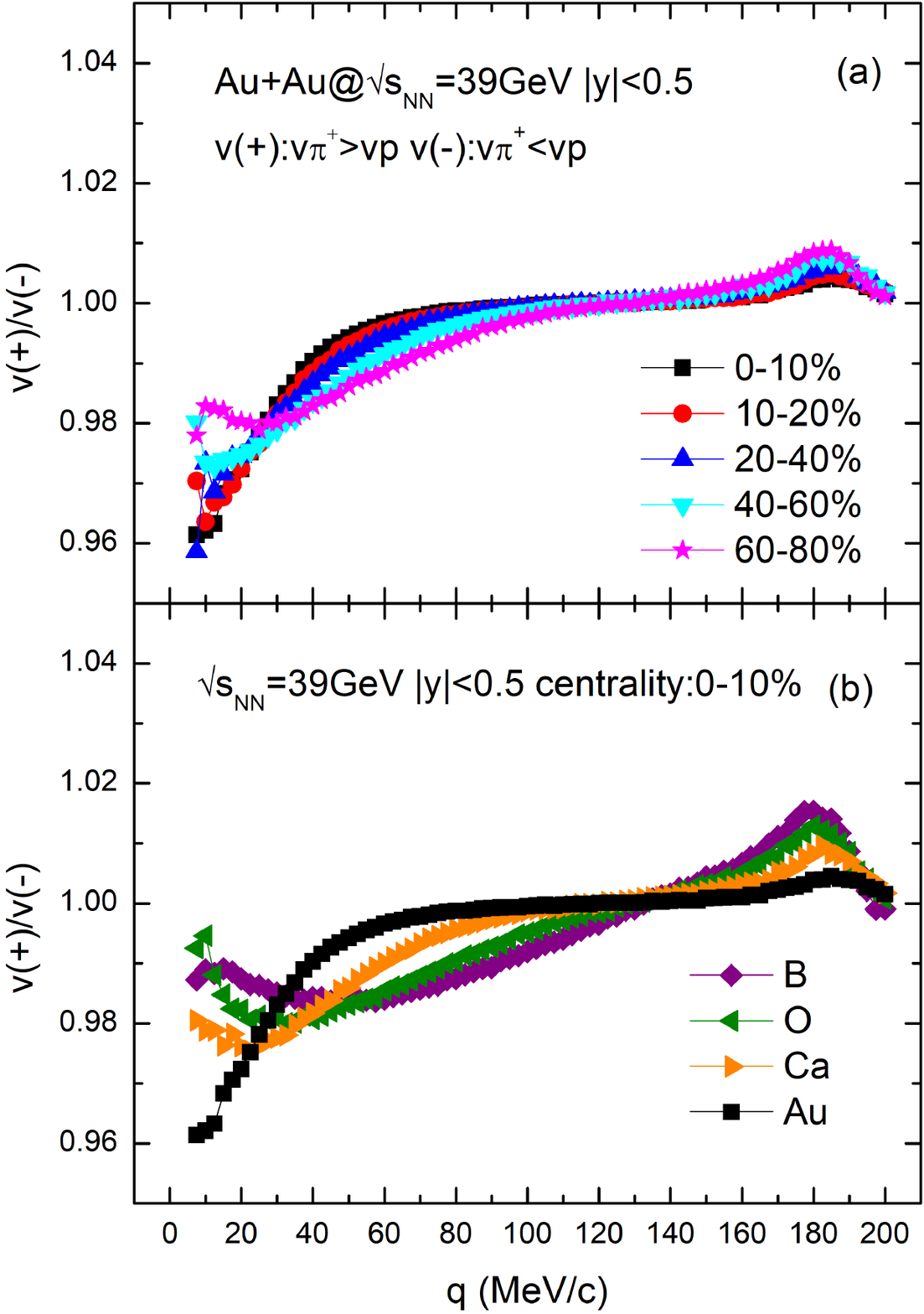}
 \centering
 \caption{Same as Fig.~\ref{fig12} but for pion-proton. 
 }
 \label{fig13}
\end{figure}

\begin{figure}[!htbp]
 \includegraphics[width=\linewidth]{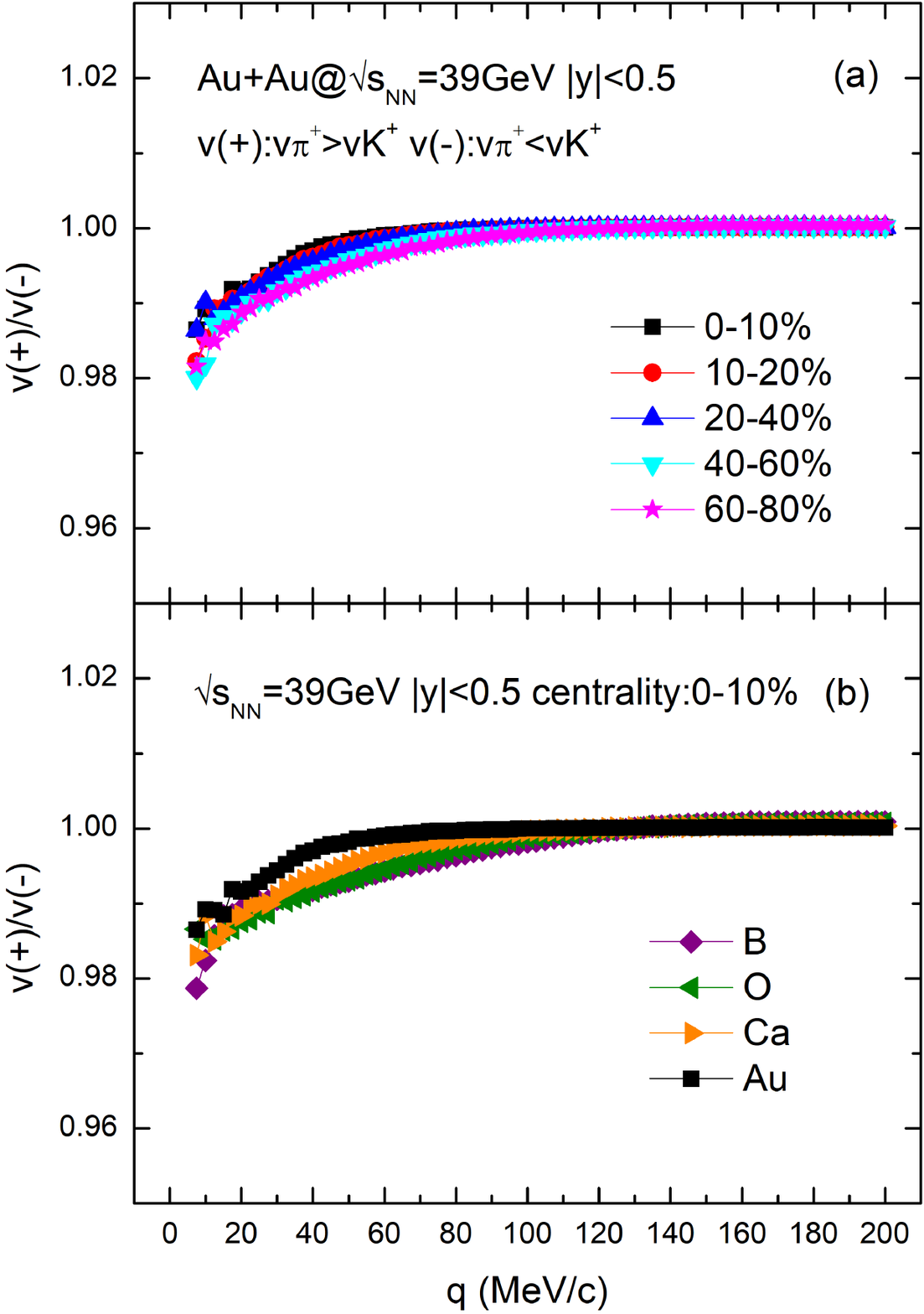}
 \centering
 \caption{Same as Fig.~\ref{fig12} but for pion-kaon. 
 }
 \label{fig14}
\end{figure}

Now we investigate centrality and system-size dependence of the like-sign (unlike-sign) nonidentical particle momentum correlation functions, such as $K^+$-$p$, $K^-$-$p$, $\pi^+$-$p$, $\pi^-$-$p$, $\pi^+$-$K^+$ and $\pi^+$-$K^-$. Fig.~\ref{fig5} (a) and (c) show results for the momentum correlation functions of $K^+$-$p$ and $K^-$-$p$ for the same centrality classes as Fig.~\ref{fig3}. The same centrality dependence is also clearly seen in Fig.~\ref{fig5} (a). 
In addition, Fig.~\ref{fig5} (b) and (d) show system-size dependence of $K^+$-$p$ and $K^-$-$p$ momentum correlation functions. 

We observe an enhanced strength of  momentum correlation function for particle pairs in smaller systems in Fig.~\ref{fig5}(b). However, centrality and system-size dependence of the unlike-sign particle momentum correlation functions are almost disappeared in Fig.~\ref{fig5}(c) and (d).
In the same way, we also investigate the effects of different centralities and system-size on the momentum correlation functions of $\pi^+$-$p$, $\pi^-$-$p$, $\pi^+$-$K^+$ and $\pi^+$-$K^-$ in Fig.~\ref{fig6} and~\ref{fig7}.  The results are similar to those in Fig.~\ref{fig5}. In Fig.~\ref{fig6} and~\ref{fig7} (b), system-size dependence of the $\pi^+$-$p$ and $\pi^+$-$K^+$ momentum correlation functions appears more sensitive to system-size only in the large system such as Au and Ca. 

Fig.~\ref{fig8} shows the dependence of radius extracted from like-sign nonidentical particle momentum correlation functions gated on centrality and system-size, where the squares, circles and triangles are results for $K^+$-$p$, $\pi^+$-$p$, and $\pi^+$-$K^+$, respectively. It is seen that the larger centrality leads to the smaller radius as shown in Fig.~\ref{fig8} (a). For system-size dependence of the radii, we find that the radii generally increase with system-size in Fig.~\ref{fig8} (b). Meanwhile, the source radii from $K^+$-$p$ correlation functions are smaller than those extracted from $\pi^+$-$p$, $\pi^+$-$K^+$ in Fig.~\ref{fig8}, which comes from stronger correlation, i.e., the smaller Bohr radius ~\cite{Kisiel2007,LEDNICKY1982,GYULASSY1979}. 
The results are similar to the previous results~\cite{Zbroszczyk2019,Siejka2019}. In an alternative viewpoint from freeze-out, the smaller $R$ of $K^+$-$p$ source indicates earlier freeze-out of $p$ and/or $K^+$ rather than $\pi^+$ and/or $K$ which has larger $R$ of $K^+$-$\pi$. In the later texts of this article, we find that the  emission order of $p$, $K$ and $\pi$ is consistent with the extracted source size.

\subsection{Velocity selected momentum correlation functions for like-sign nonidentical particles}

Momentum correlation functions of unlike particles can provide an independent constrain on their mean emission order by simply making velocity selections~\cite{Gelderloos1994,Gelderloos1995,DGourio2000,Lednicky1996,HuangBS1}. The physics details can be seen in the literature~\cite{wtt2023}, which has been well applied to explore the light (anti)nuclei momentum correlation functions. 
 Figs.~\ref{fig9},~\ref{fig10} and~\ref{fig11} present the velocity-gated momentum correlation functions as well as velocity difference ($\Delta$v) spectra of like-sign unlike particles pairs $K^+$-$p$, $\pi^+$-$p$ and $\pi^+$-$K^+$ for 39 GeV $_{79}^{197}\textrm{Au}+_{79}^{197}\textrm{Au}$ collisions at different centralities of $0-10$ $\%$, $10-20$ $\%$, $20-40$ $\%$, $40-60$ $\%$, and $60-80$ $\%$, respectively. 

Fig.~\ref{fig9},~\ref{fig10} and~\ref{fig11} (a), (b) show centrality dependence of velocity-gated momentum correlation functions and velocity difference ($\Delta$v) spectra of $K^+$-$p$, $\pi^+$-$p$ and $\pi^+$-$K^+$ pairs, respectively. For $K^+$-$p$ and $\pi^+$-$p$ pairs, the momentum correlation functions with $v_{K^+}$ $>$ $v_{p}$ ($v_{\pi^+}$ $>$ $v_{p}$) are stronger than the ones with the reverse situation $v_{K^+}$ $<$ $v_{p}$ ($v_{\pi^+}$ $<$ $v_{p}$) in Fig.~\ref{fig9} and~\ref{fig10} (a). The comparison of  two velocity-gated correlation strengths gives that the mean order of emission of protons  are  emitted averagely earlier than $\pi^+$ and $K^+$ according to the criteria~\cite{wtt2023}. The similar trend for $\pi^+$-$K^+$ pairs is not so obvious overall in Fig.~\ref{fig11} (a).

Meanwhile, Fig.~\ref{fig9},~\ref{fig10} and~\ref{fig11} (b) present  velocity difference spectra for $K^+$-$p$, $\pi^+$-$p$ and $\pi^+$-$K^+$ pairs, respectively. The velocity difference spectra  are all asymmetric due to the mean emission order. In addition, an enhanced difference between the momentum correlation functions for $K^+$-$p$ ($\pi^+$-$p$ or $\pi^+$-$K^+$) pairs with $v_{p} > v_{K^+}$ ($v_{p} > v_{\pi^+}$ or $v_{\pi^+} > v_{K^+}$) and ones on the reverse situation at larger centrality, which manifests larger interval of the mean emission order for unlike particles in peripheral collisions. 

Their ratios shown in Fig.~\ref{fig12},~\ref{fig13} and~\ref{fig14} (a) also illustrate the above phenomenon. The system-size dependence for $K^+$-$p$, $\pi^+$-$p$ and $\pi^+$-$K^+$ pairs is found by the fact that momentum correlation functions with $v_{K^+}$ $>$ $v_{p}$ ($v_{\pi^+}$ $>$ $v_{p}$) are stronger than the ones with the reverse situation $v_{K^+}$ $<$ $v_{p}$ ($v_{\pi^+}$ $<$ $v_{p}$) in Fig.~\ref{fig9},~\ref{fig10} and~\ref{fig11} (c). Correspondingly, the velocity difference spectra for $K^+$-$p$, $\pi^+$-$p$ and $\pi^+$-$K^+$ pairs are all asymmetric about $\Delta$v = 0 caused by the average emission order in Fig.~\ref{fig9},~\ref{fig10} and~\ref{fig11} (d).
All the correlation ratios for the pairs of $K^+$-$p$, $\pi^+$-$p$ and $\pi^+$-$K^+$ generally  demonstrate a dip below 1 in panel (a) of  
Fig.~\ref{fig12},~\ref{fig13} and~\ref{fig14}, which indicate that the emission order is, respectively, $\tau_{K^+} > \tau_p$, $\tau_{\pi^+} > \tau_p$, and $\tau_{\pi^+} > \tau_{K^+}$. Then we can get $\tau_p < \tau_{K^+} < \tau_{\pi^+}$, i.e. the proton is on average emitted earliest, then $K^+$ is in the middle, and $\pi^+$ is the latest. This conclusion seems to consistent with the different freeze-out time for $\pi$, $K$ and $p$ in  a blast wave model analysis \cite{LiuFH}, where the freeze-out temperature increases with the particle$^{,}$s mass ($\pi$, K, and proton), but the kinetic freeze-out volume decreases with the increase of particle mass, i.e. it indicates a mass-dependent differential kinetic freeze-out scenario. As mentioned above, the source size extracted from different meson-baryon or meson-meson pairs in this work (Fig.~\ref{fig8}) is generally  consistent with the mass-dependent kinetic freeze-out volume.  
However, the emission chronology among $\pi$, $K$ and $p$ looks a reverse mass-dependent in comparison with the emission order for light nuclei, such as $p$, $d$ and $t$ as investigated in Ref.~\cite{wtt2023}, in which 
the heavier species  are emitted later in the small relative momentum region due to the nucleonic coalescence picture. 

In addition, the system-size dependence of velocity-gated momentum correlation functions is also generally observed by their ratios in Fig.~\ref{fig12},~\ref{fig13} and~\ref{fig14} (b). 
With the decreasing of system-size, we  observe an enhanced difference between the momentum correlation functions for $K^+$-$p$ ($\pi^+$-$p$ or $\pi^+$-$K^+$) pair with $v_{p} > v_{K^+}$ ($v_{p} > v_{\pi^+}$ or $v_{\pi^+} > v_{K^+}$) and the ones with the reverse situation in Fig.~\ref{fig9},~\ref{fig10} and~\ref{fig11} (c).  The dependence on system-size for the momentum correlation functions of meson-baryon or meson-meson particle pairs is caused by the emission source of different particles as shown in Fig.~\ref{fig8}(b). In the viewpoint of emission order, we can generally  say that the emission duration among $p$, $K$ and $\pi$ could be larger for the smaller system than the larger system, especially for $\pi$-hadron correlations (Fig. 13 and 14). For $K^+$-p combination (Fig. 12), it is noted that the ratio shows first drop to a dip, then rises for light collision systems, which is different from the case of Au + Au collision. In particular, the ratios display the values above 1 in the lowest relative momentum, which illustrates that $\tau_p$ $>$ $\tau_K^+$,  i.e. proton is emitted on average earlier than $K^+$. At around $q \sim$ 20 MeV/c, the ratios display the values below 1, i.e.  $\tau_p$ $<$ $\tau_K^+$, then $\tau_p$ $\simeq$ $\tau_K^+$ at  larger $q$. However, the fine structure for each ratios of meson-meson and meson-baryon correlations demonstrates slight differences, which indicates slightly different emission chronology.
 
\section{SUMMARY}

In summary, with the AMPT model complemented by the Lednick$\acute{y}$ and Lyuboshitz analytical method, we have constructed and analyzed the momentum correlation functions of like-sign (unlike-sign) particle for heavy-ion collisions with different system sizes and centralities for $\sqrt{s_{NN}}$ = 39 GeV Au + Au collisions. We present a comparison of like-sign (unlike-sign) $K$-$p$, $\pi$-$K$ and $\pi$-$p$ momentum correlation functions with the experimental data from the RHIC-STAR collaboration~\cite{Zbroszczyk2019,Siejka2019}. Taking the same transverse momentum and rapidity phase space coverage corresponding to the experimental situation as well as the maximum hadronic rescattering time selected by mesons of 400 $fm/c$ and proton of  700 $fm/c$ in AMPT, it is found that the like-sign (unlike-sign) $K$-$p$, $\pi$-$K$ and $\pi$-$p$ momentum correlation functions simulated by the present model can match the experimental data. We further study centrality and system-size dependence of momentum correlation functions for identical and nonidentical particle pairs, respectively, which is in the condition of the maximum hadronic rescattering time of 100 $fm/c$  in AMPT. The shape of momentum correlation functions for particle pairs is consistent with previous works~\cite{zzq2014,Adamczewski2020,Star-prc2005}, 
which is caused by both QS and FSI. 

The centrality dependence of momentum correlation functions for particles is investigated by $_{79}^{197}\textrm{Au}+_{79}^{197}\textrm{Au}$ collisions at  five centralities of $0-10$ $\%$, $10-20$ $\%$, $20-40$ $\%$, $40-60$ $\%$, and $60-80$ $\%$ at $\sqrt{s_{NN}}$ = 39 GeV. It is found that with increasing centralities from center to periphery, the momentum correlation functions for particles become stronger, which is consistent with the   emission from the smaller source. The momentum correlation functions of particles are  sensitive to system-size through studying $_{5}^{10}\textrm{B}+_{5}^{10}\textrm{B}$, $_{8}^{16}\textrm{O}+_{8}^{16}\textrm{O}$, $_{20}^{40}\textrm{Ca}+_{20}^{40}\textrm{Ca}$ and $_{79}^{197}\textrm{Au}+_{79}^{197}\textrm{Au}$ in central collisions, and used to obtain the emission source-size of particles which is self-consistent with their system-size. Furthermore, momentum correlation functions between nonidentical particles shed light on important information about the average emission sequence of them. Through the correlation functions gated with the velocity, it is  deduced that protons are generally emitted earliest, $K$ is in middle, and $\pi$ is the latest in the small relative momentum region, which indicates mass-dependent kinetic freeze-out scenario. Experimental analysis along this direction is expected. 

\section*{Acknowledgments}
This work was supported in part by the National Natural Science Foundation of China under contract Nos.  11890710, 11890714, 11875066, 11925502, 11961141003, 11935001, 12147101 and 12047514,  the Strategic Priority Research Program of CAS under Grant No. XDB34000000, National Key R\&D Program of China under Grant No. 2018YFE0104600 and 2016YFE0100900,  Guangdong Major Project of Basic and Applied Basic Research No. 2020B0301030008, and the China PostDoctoral Science Foundation under Grant No. 2020M681140.


\bibliography{hbtpikp}

\end{document}